\documentclass[11pt]{article}
\usepackage{amssymb,amsmath,graphicx,epsfig,subfig,color,hyperref,url}
\usepackage[margin=1in]{geometry}
\usepackage[retainorgcmds]{IEEEtrantools}
\usepackage{slashed}
\usepackage{multirow}
\usepackage{epstopdf}
\usepackage{caption}
\usepackage{url}
\usepackage{float}
\usepackage{cite}
\usepackage{caption}
\usepackage{mathptmx}
\usepackage{wrapfig}
\usepackage{epsfig}
\usepackage{graphicx}
\usepackage{footnote}
\usepackage[utf8]{inputenc}

\begin{document}
\tolerance=100000
\thispagestyle{empty}
\setcounter{page}{0}

\def\cO#1{{\cal{O}}\left(#1\right)}
\newcommand{\nn}{\nonumber}
\newcommand{\be}{\begin{equation}}
\newcommand{\ee}{\end{equation}}
\newcommand{\br}{\begin{eqnarray}}
\newcommand{\er}{\end{eqnarray}}
\newcommand{\ba}{\begin{array}}
\newcommand{\ea}{\end{array}}
\newcommand{\bi}{\begin{itemize}}
\newcommand{\ei}{\end{itemize}}
\newcommand{\bn}{\begin{enumerate}}
\newcommand{\en}{\end{enumerate}}
\newcommand{\bc}{\begin{center}}
\newcommand{\ec}{\end{center}}
\newcommand{\ul}{\underline}
\newcommand{\ol}{\overline}
\newcommand{\ra}{\rightarrow}
\newcommand{\sm}{${\cal {SM}}$}
\newcommand{\as}{\alpha_s}
\newcommand{\aem}{\alpha_{em}}
\newcommand{\ycut}{y_{\mathrm{cut}}}
\newcommand{\susy}{{{SUSY}}}
\newcommand{\Dir}{\kern -6.4pt\Big{/}}
\newcommand{\Dirin}{\kern -10.4pt\Big{/}\kern 4.4pt}
\newcommand{\DDir}{\kern -10.6pt\Big{/}}
\newcommand{\DGir}{\kern -6.0pt\Big{/}}
\def\Ecm{\ifmmode{E_{\mathrm{cm}}}\else{$E_{\mathrm{cm}}$}\fi}
\def\lsim{\buildrel{\scriptscriptstyle <}\over{\scriptscriptstyle\sim}}
\def\gsim{\buildrel{\scriptscriptstyle >}\over{\scriptscriptstyle\sim}}
\def\jp #1 #2 #3 {{J.~Phys.} {#1} (#2) #3}
\def\pl #1 #2 #3 {{Phys.~Lett.} {#1} (#2) #3}
\def\np #1 #2 #3 {{Nucl.~Phys.} {#1} (#2) #3}
\def\zp #1 #2 #3 {{Z.~Phys.} {#1} (#2) #3}
\def\pr #1 #2 #3 {{Phys.~Rev.} {#1} (#2) #3}
\def\prep #1 #2 #3 {{Phys.~Rep.} {#1} (#2) #3}
\def\prl #1 #2 #3 {{Phys.~Rev.~Lett.} {#1} (#2) #3}
\def\mpl #1 #2 #3 {{Mod.~Phys.~Lett.} {#1} (#2) #3}
\def\rmp #1 #2 #3 {{Rev. Mod. Phys.} {#1} (#2) #3}
\def\sjnp #1 #2 #3 {{Sov. J. Nucl. Phys.} {#1} (#2) #3}
\def\cpc #1 #2 #3 {{Comp. Phys. Comm.} {#1} (#2) #3}
\def\xx #1 #2 #3 {{#1}, (#2) #3}
\def\NP(#1,#2,#3){Nucl.\ Phys.\ \issue(#1,#2,#3)}
\def\PL(#1,#2,#3){Phys.\ Lett.\ \issue(#1,#2,#3)}
\def\PRD(#1,#2,#3){Phys.\ Rev.\ D \issue(#1,#2,#3)}
\def\preprint{{preprint}}
\def\Ord{\lower .7ex\hbox{$\;\stackrel{\textstyle <}{\sim}\;$}}
\def\OOrd{\lower .7ex\hbox{$\;\stackrel{\textstyle >}{\sim}\;$}}
\def \ptz{\rm p_T(Z)}  
\def \invfb{fb^{-1}}
\def\bul{\bullet}
\def\lapp{\mathrel{\rlap{\raise.5ex\hbox{$<$}}
                    {\lower.5ex\hbox{$\sim$}}}}
\def\gapp{\mathrel{\rlap{\raise.5ex\hbox{$>$}}
                    {\lower.5ex\hbox{$\sim$}}}}
\begin{flushright}
\end{flushright}
\begin{center}
{\Large \bf
A QCD analysis of CMS inclusive differential 
Z production data at $\sqrt{s}$ = 8 TeV} 
\\[1.00cm]
\end{center}
\begin{center}
{\large Rajdeep M Chatterjee$^a$, Monoranjan Guchait$^a$, 
Ringail\. e Pla\v cakyt\. e$^b$  
}\\[0.3 cm]
{\it 
\vspace{0.2cm}
$^a$ Department of High Energy Physics\\
Tata Institute of Fundamental Research\\ 
Homi Bhabha Road, Mumbai-400005, India.\\
\vspace{0.2cm}
$^b$ Deutsches Elektronen-Synchrotron DESY \\
Notkestr. 85, D-22607, Hamburg, Germany
}
\end{center}

\vspace{2.cm}

\begin{abstract}
{\noindent\normalsize 
The parton distribution functions (PDFs) of the proton 
are one of the essential ingredients to describe 
physics processes at hadron colliders. 
The Z boson production data at the LHC have a potential to constrain
PDFs, especially the gluon distribution. 
In this study the CMS measurement of the inclusive double differential 
Z boson production cross section in terms of transverse momentum and rapidity
are compared to the next-to-leading order theory predictions
at the center of mass energy, $\sqrt{s}$=8 TeV with an integrated 
luminosity of 19.71 $\invfb$. In addition, the sensitivity of 
this measurement to PDFs is studied within the framework of the HERAFitter.
A moderate improvement to the gluon distribution is observed at the 
Bjorken $x \approx 0.1$ region. However, in order to obtain 
further improvement to the gluon distribution in 
the global fits, the higher-order theory calculations 
accessible via fast techniques are necessary. 
}
\end{abstract}
\vspace{2cm}
\hskip1.0cm
\newpage
\section{Introduction}
The precision measurements of the Standard Model (SM) of particle physics are 
one of the top priority programmes at the LHC.
The accurate theory predictions are necessary in order to completely 
exploit the
potential of the SM measurements.
Importantly, in the hadron colliders parton distribution 
functions (PDFs) are one of the necessary ingredients for theory 
predictions.
Naively, the PDF $f_i(x,Q^2)$ represent the probability of finding 
a parton of flavour $i$ (where $i$: g(gluons),q (quarks); q=u,d,c,s..)
inside a proton carrying a fraction $x$ of the momentum of the proton 
at the scale $Q$, called the factorization scale related with the 
hard scale of the involved physical process.
The PDFs cannot be derived from the 
first principles of Quantum Chromodynamics (QCD)~\cite{Campbell:2006wx}
and have to be constrained experimentally. 
The PDFs are constrained primarily by the Deep Inelastic Scattering (DIS) 
data. Additional constraints come from the Fixed-target, 
Tevatron and LHC measurements [for more details, see for example, 
the reviews in Ref.\cite{Martin:2009iq,Rojo:2015acz} and references therein].

Currently, various published SM measurements at the LHC with the center 
of mass energies 7 and 8 TeV are already used in the global 
PDF 
fits~\cite{Harland-Lang:2014zoa,Dulat:2015mca,Rojo:2015nxa,Alekhin:2013nda}.
Besides these global PDF fitting efforts,
the sensitivity of the particular LHC measurement to PDFs are 
also studied. For example, PDFs are constrained using the measurement 
of the W and Z production 
to strange quark distribution 
in ATLAS~\cite{Aad:2012sb}, the CMS W charge 
asymmetry~\cite{Chatrchyan:2013mza} and W in association with the 
charm quark measurements at 7 TeV~\cite{Chatrchyan:2013uja},
W boson production in association with a single charm quark 
in ATLAS~\cite{Aad:2014xca}, the inclusive jet cross sections 
from the LHC~\cite{Watt:2013oha,Aad:2013lpa,Khachatryan:2014waa,Rojo:2014kta},
top quark pair production~\cite{Czakon:2013tha,Guzzi:2014wia} etc.
In this current analysis, the impact to the PDFs 
of the CMS production cross section
measurement of the Z boson decaying to a pair of 
muons~\cite{CMS-PAS-SMP-13-013,Khachatryan:2015oaa} is studied. 
This measurement is performed in various bins of transverse momentum ($\ptz$)
and rapidity ($Y(Z)$) of the Z boson. 
The QCD analysis is performed at the Next-to-Leading order(NLO)
in the framework of the HERAFitter~\cite{Alekhin:2014irh}. 
Along with the CMS Z boson measurements, the inclusive
HERA-I DIS~\cite{Aaron:2009aa} and the 
CMS W muon asymmetry data~\cite{Chatrchyan:2013mza} are used in this study.

This note is organized as follows. In Sec.2 we discuss the inclusive Z boson
production at the LHC while in Sec.3 the correlation 
studies of the different partons with the Z boson production are 
presented. The general settings of the QCD analysis and the results are  
presented in Sec.4 and 5 respectively followed by a summary in Sec.6.

\section{Inclusive Z production at LHC}
\label{Zpt}
At the LHC experiment, the Z boson production cross section
is one of the high priority measurements to understand the detector 
performance as well as for precision test$\textbf{s}$ of the SM.
The inclusive Z boson production at the LHC is initiated by 
the sub-processes,
\br
\rm {q \bar q, \ gq  \to Z+n~jets}
\label{eq:zprod}   
\er
at the leading order (LO), where ${\rm n} \ge 0$. 
At the low $\rm p_T (Z)$ region, 
$\ptz \lsim 20$~GeV, the sub-process with initial state 
$\rm {q \bar q}$ has significant contribution to the total cross section,
while in the high $\ptz$ region,
$\ptz \gsim {\rm M_Z/2}$, the sub-process $\rm gq$ becomes dominant
($\sim$ 70 - 80\%).
These relative sub-process contributions are also valid at the NLO 
accuracy~\cite{Watt:2013oha,Malik:2013kba,Brandt:2013hoa}.
Therefore, the inclusive Z boson production cross section 
is one of the potential measurement to 
probe the gluon density inside the proton at the 
LHC~\cite{Bandurin:2004ch}.

The vector boson(W,Z) production in hadron colliders has a very rich 
physics potential and is well studied in the 
literature~\cite{Altarelli:1979ub,Altarelli:1984pt,Collins:1984kg}.
Currently, the inclusive Z boson production in its leptonic decay 
channel is computed at the next-to-next-to leading order(NNLO) 
level(${\cal O}(\alpha_s^2)$) 
by several groups~\cite{Melnikov:2006kv,Catani:2009sm} and 
the K-factor which is defined to be the ratio of the NNLO and LO 
cross sections, is estimated to be about $\sim 1.4-1.6$ 
depending on the kinematic phase space. These NNLO predictions are 
equivalent to NLO for the Z+n-jet 
(n$ \ge 1$) 
process which is of the same order(${\cal O}(\alpha_s^2)$). 
It is to be noted that these NLO predictions work quite well 
for the high $\ptz$ regime. However, for the low $\ptz$
range where the soft gluon emission with very 
low transverse momentum takes place, the NLO predictions 
may become unstable.
In this low $\ptz$ region, the fixed order perturbative 
calculation fails due to the presence of large logarithmic terms 
$\sim {\rm log(M_Z/p_T)}$. Therefore, in order to make the 
calculation realistic at this low transverse momentum region, 
the large logarithms must be re-summed 
to all orders of $\alpha_s$.    
In Ref.~\cite{Balazs:1997xd,Ladinsky:1993zn} 
the soft gluon non perturbative effects are taken into account 
re-summing all logarithm terms in the Z boson
production and it can be computed using the software package {\tt ResBos}.

The full NNLO QCD corrections 
(${\cal O}(\alpha_s^2)$) to the ${\rm p_T(Z)}$ and ${\rm p_T(W)}$ 
distributions in association with jets 
became available very 
recently~\cite{Boughezal:2015ded, Boughezal:2015dva,Boughezal:2016isb,Andersen:2016vkp}. 
However, before that  
an attempt has been made to compute the two loop QCD 
corrections to the process ${\rm gg \to Zj}$ using the helicity 
amplitudes~\cite{Gehrmann:2013vga} method. 

At the LHC, both the CMS and the ATLAS collaborations have 
measured the inclusive Z boson 
production cross section in proton-proton collisions at 
the center of mass 
energies of $\sqrt{s}$=7~\cite{Aad:2014xaa} and 
8 TeV~\cite{Khachatryan:2015oaa}.  
These measurements are 
performed by identifying the Z boson in both the electron and muon 
channels and  the
results are presented in terms of the differential distributions of 
the Z boson.
In particular, the CMS 
measurement relevant in the present context,
the differential cross sections are obtained  
in terms of the $\ptz$ 
and absolute $Y(Z)$ of the Z 
boson~\cite{CMS-PAS-SMP-13-013,Khachatryan:2015oaa}
based on the data sample of pp collisions at $\sqrt{s} = 8$ TeV
corresponding to an integrated luminosity of 19.71 $\invfb$.
The results are presented in five absolute rapidity bins 
ranging from 0-2.0 and the entire $\ptz$ range has been divided into 10 bins
reaching up to 1 TeV. 
The main goal of this current study is to test the sensitivity of the 
high $\ptz$ data
to the gluon PDF, hence the measurements of the first two 
bins (0-20 GeV and 20-40 GeV) are not included in the QCD analysis.
The various sources of uncertainties related 
to the measurement techniques and the  
background estimation are 
obtained with the bin-to-bin correlation for each uncertainty 
source and accounted for
in terms of covariance matrices.   
It is observed that the luminosity measurement is a 
main source of uncertainty amounting to 2.6\% leading a 
total uncertainty $\sim$3-4\% in the 
measurement~\cite{CMS-PAS-SMP-13-013,Khachatryan:2015oaa}.
It is important to note that in this measurement no selection has been
made on jets accompanied with the Z boson in order to avoid  
possibly a sizable (5-10\%) contribution 
to the total systematic uncertainties from jet energy scale measurements.
The measured cross sections are 
presented unfolding the detector effects at the parton level 
to be compared with higher order theory 
predictions. In the Ref~\cite{Khachatryan:2015oaa}, a 
comparison of data to the theoretical predictions from the {\tt FEWZ} 
computation~\cite{Li:2012wna} for all rapidity bins
is presented. The level of agreement between the data and the 
theoretical prediction is found to be
of the  ${\cal O}(\sim$10\%)  across all $\ptz$  and $Y(Z)$ bins, 
which is within the 
uncertainties of the measurement and the theoretical prediction. 
The uncertainty 
in the theoretical predictions includes the variation of the 
QCD scales and PDF. 

In the present QCD analysis, the inclusive Z boson differential
cross sections are calculated at NLO using the 
{\tt MCFM}~\cite{Campbell:2010ff}
interfaced to {\tt ApplGrid}~\cite{Carli:2010rw}.
The transverse$\textbf{e}$ momentum of the 
leading (sub-leading) muons are required to be 
greater than 25(10) GeV and less than 
2.4(2.1) in absolute rapidity, whereas the dimuon invariant mass is 
selected to be
within 81-101 GeV.     
The NLO CT10 PDF set~\cite{Lai:2010vv} is used for the Z-boson event 
generation and the factorization ($\mu_F$) and 
the renormalization ($\mu_R$) scales are set to the dynamical 
scale $\mu_0 = \sqrt{(\ptz)^2 + M_Z^2}$. 
The uncertainties in the $\ptz$ distribution  
due to the choices of QCD scales and PDF are also computed.
The uncertainty due to the QCD scales is obtained
by varying the scales, $\mu_R,\mu_F$= \{ 1/2, 2\}$\times \mu_0$.
It is observed that the QCD scale uncertainty 
is a size of about $\sim$5\% at the low $\ptz$
region and rises to $\sim$ 7\% at the high $p_T(Z)$ region.
The PDF uncertainty is derived following the asymmetric uncertainty
prescription~\cite{Lai:2010nw} 
by generating the $p_T(Z)$ spectrum 
for all the
up and down type eigenvectors of the NNPDF2.3~\cite{Ball:2014uwa} PDF set 
and found to be about 2-3\%. Evidently, 
the scale uncertainty dominates over the other theoretical uncertainties.  

\begin{figure}[ht]
  \centering
  \includegraphics[scale=0.40]{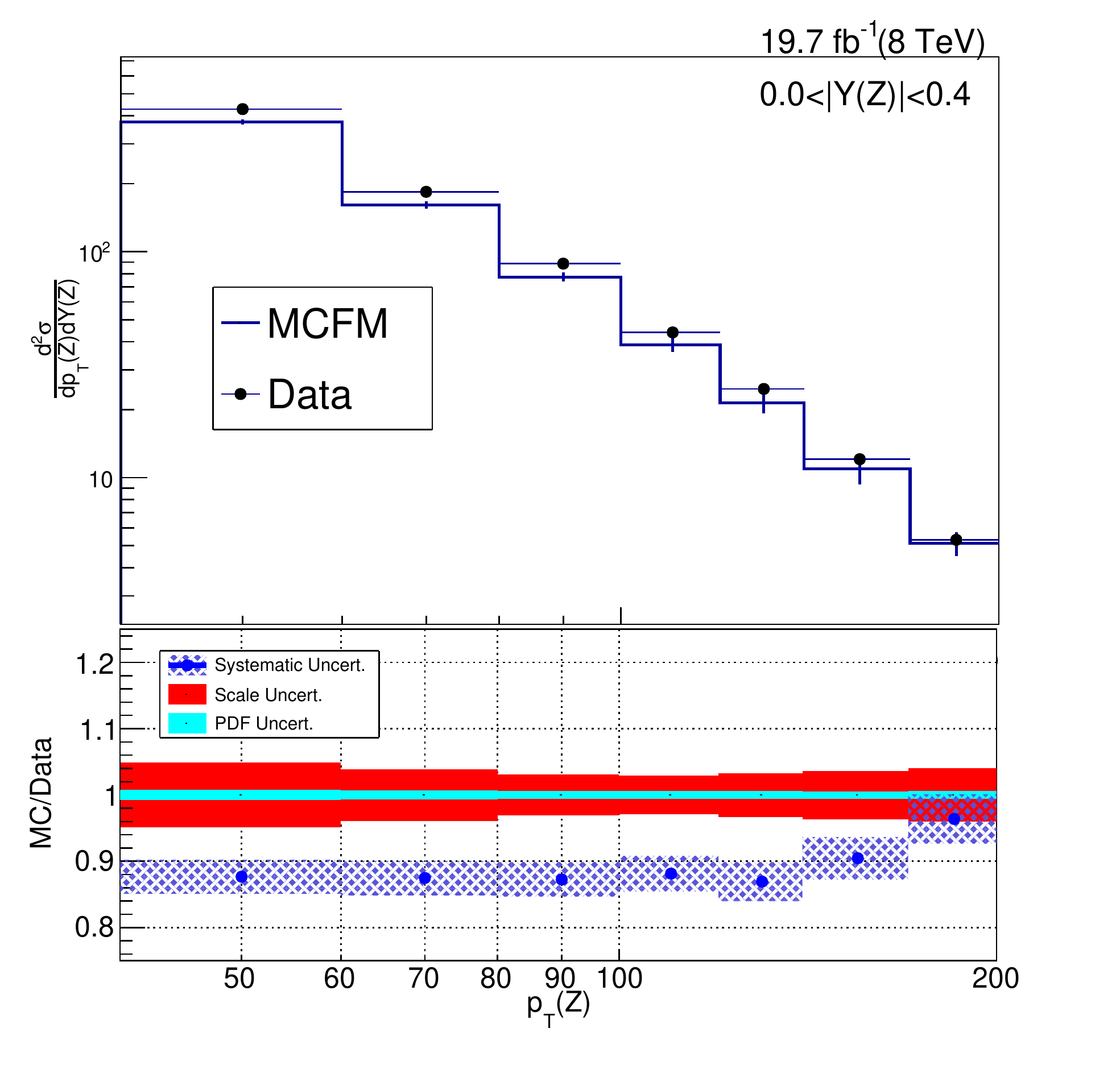}
  \includegraphics[scale=0.40]{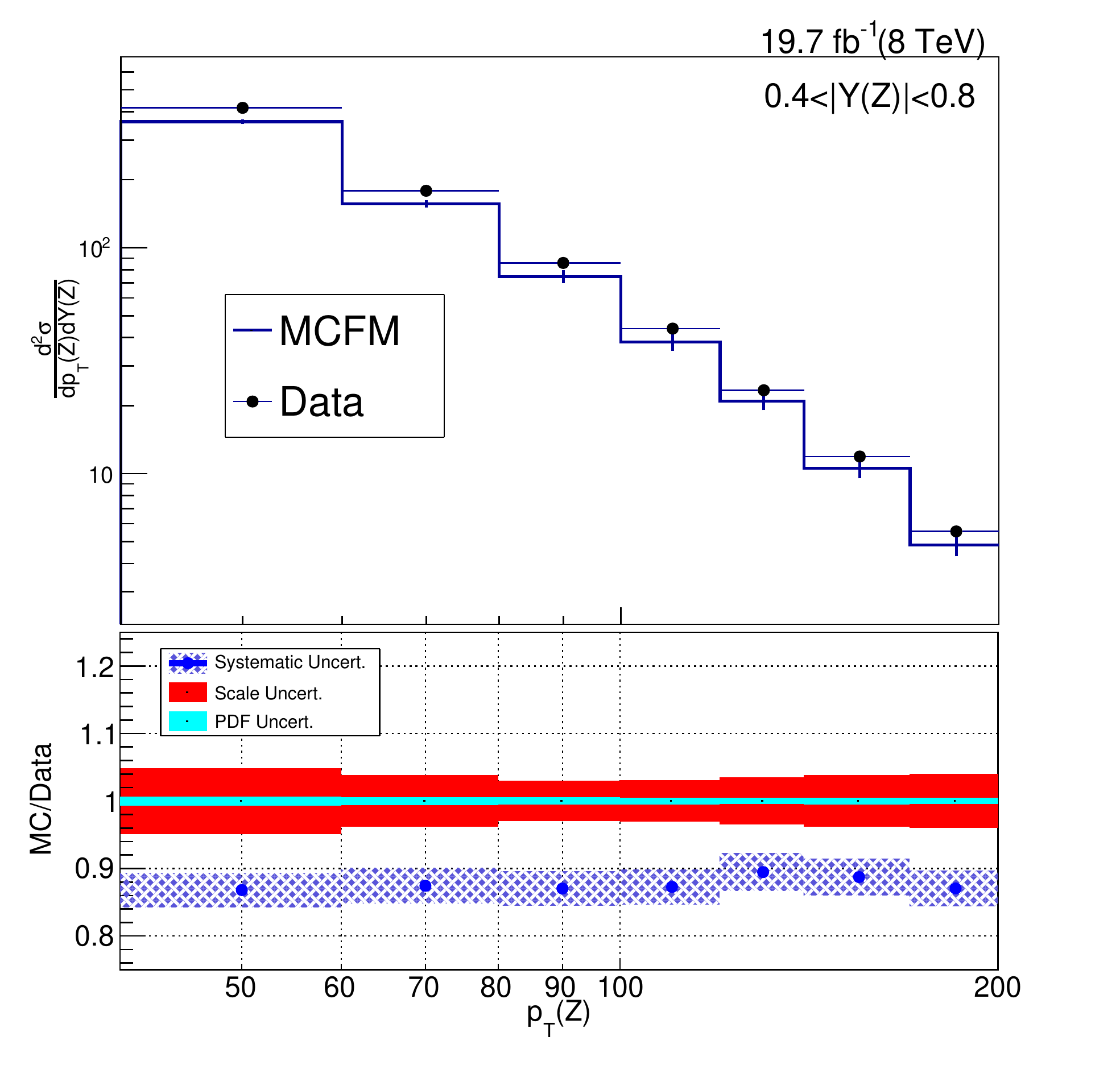}
  \includegraphics[scale=0.40]{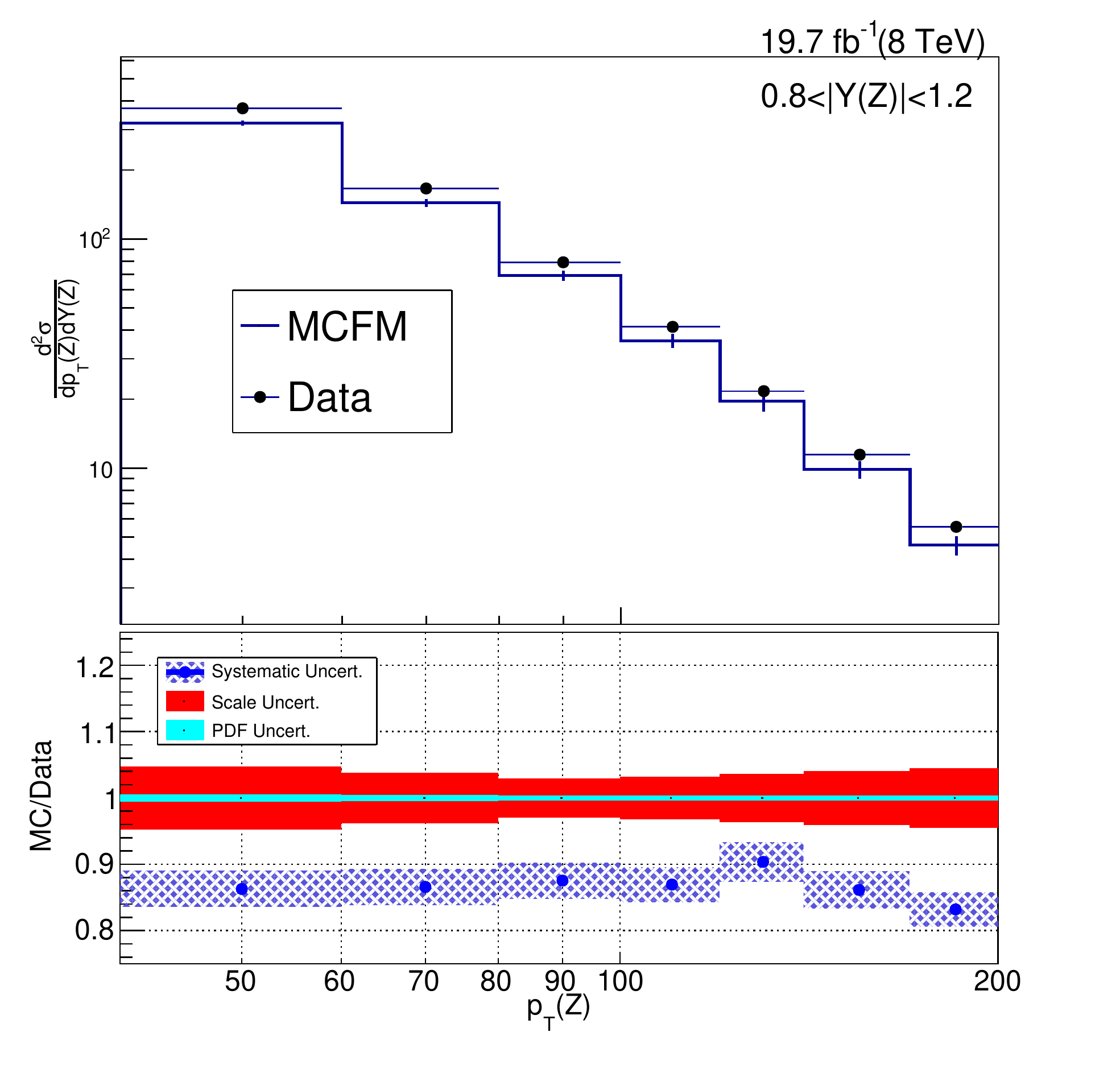}
  \includegraphics[scale=0.40]{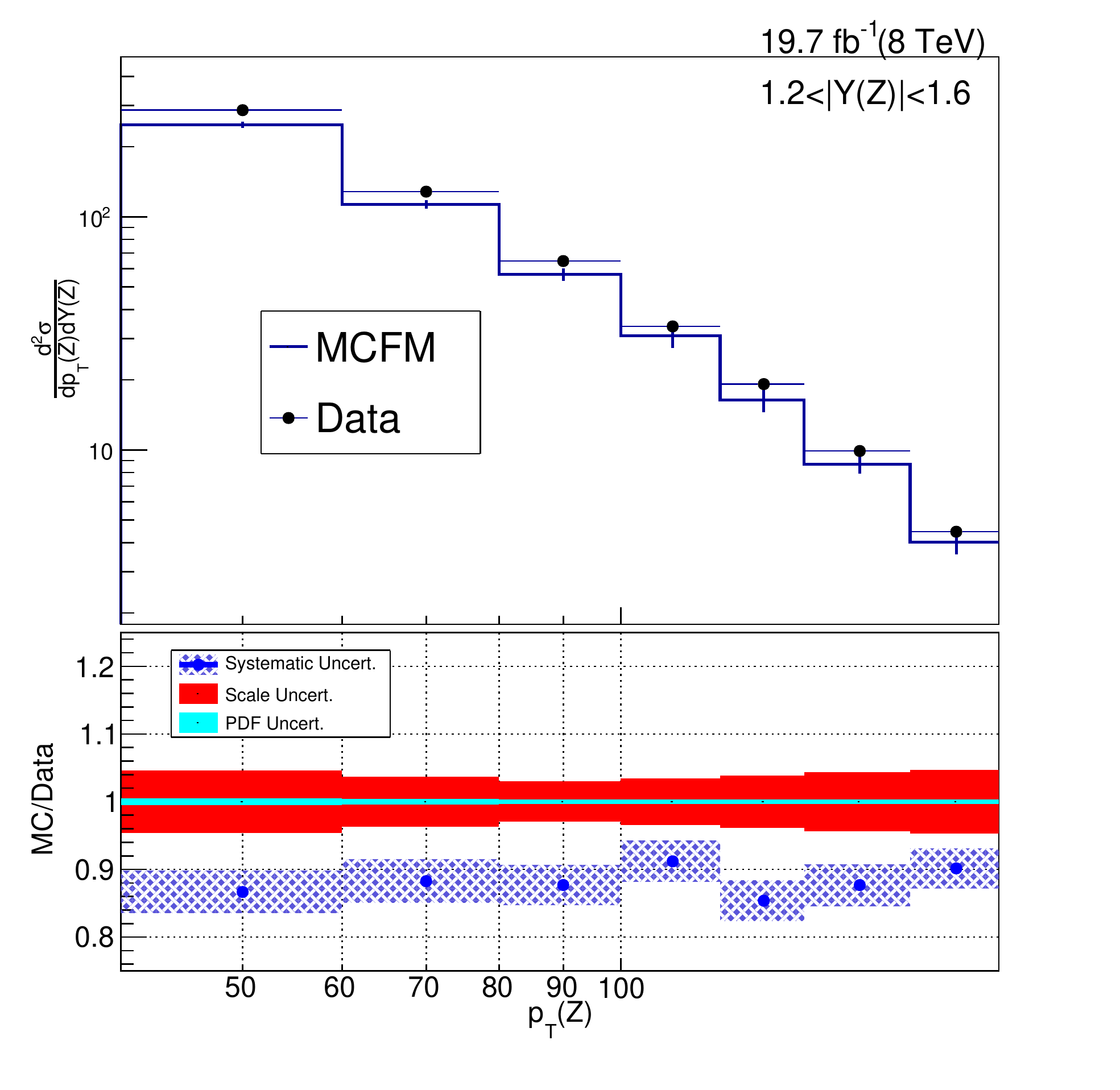}
  \includegraphics[scale=0.40]{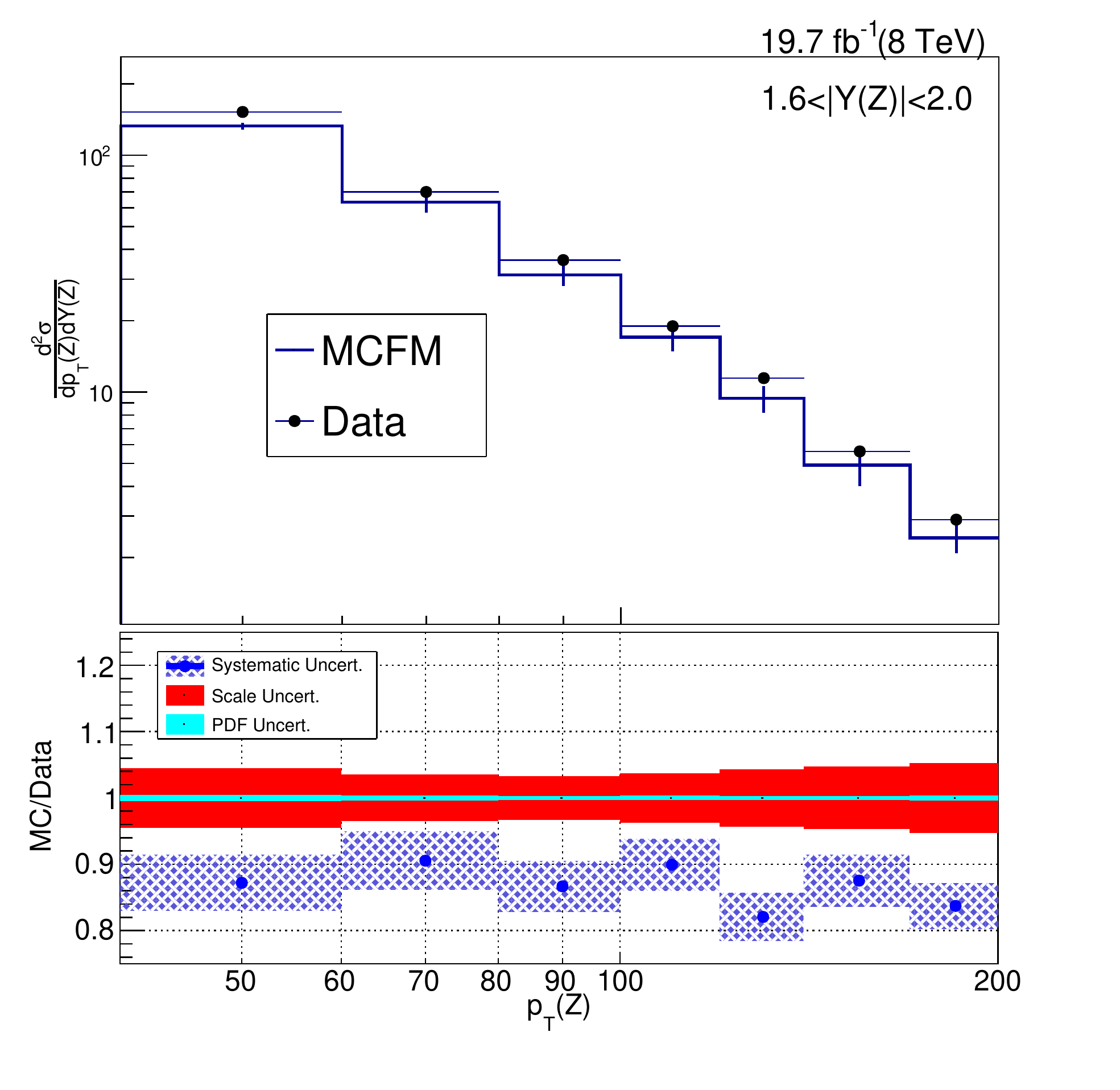}
  \caption{
The comparison of Z boson CMS data to the theory predictions obtained 
from the MCFM~\cite{Campbell:2010ff} 
in different Z rapidity (Y(Z)) bins.
}
\label{fig:datath}
\end{figure}

The comparison between the MCFM theory predictions and the 
measured double differential Z boson cross section 
normalized to the 
inclusive cross section 
for various $\ptz$ and $Y(Z)$ bins
corresponding to an integrated luminosity of
19.71$\invfb$ is presented in Fig.\ref{fig:datath}.
In addition, the ratios of the data 
and the theory predictions are also presented 
along with the PDF and QCD scale uncertainty band.
As can be seen in Fig.~\ref{fig:datath}, a disagreement of about 10\% exists 
between the measured cross section and the MCFM theory prediction across 
all the rapidity bins except for the last bin($1.6<Y(Z)<2.0$)
where the discrepancy is even larger.
Note that a similar level of agreement of the inclusive 
Z boson data is also observed with the theory predictions provided by     
FEWZ~\cite{Khachatryan:2015oaa}. Recently, in the 
Ref.~\cite{Boughezal:2016yfp}, the ATLAS and CMS data at 7 TeV 
for V+jet(V=W,Z) processes are compared with NNLO theory predictions.   
\section{Parton Correlation}
\label{Corr}
In order to understand the sensitivity of the initial PDFs
to the Z boson production cross section, the correlation of  
the corresponding PDFs with the cross section
is studied using the NNPDF2.3~\cite{Ball:2014uwa} PDF set.
The correlation function,
${\rm Q}_i$ for each $i$-th parton($i$=g,u,d,s) is computed
by evaluating means and standard deviations from the set of $N_{\rm rep}$
as,
\br
{\rm Q}_i[\sigma_{incl}, xf_i(x,\mu^2)]&=& 
\frac{N_{\rm rep}}{N_{\rm rep}-1}{{\cal F}_i}(\sigma_{\rm incl},x,\mu^2)\nn \\ 
{\rm where}\nn \\
{{\cal F}_i}(\sigma_{\rm incl},x,\mu^2)&=&
\frac{<\sigma_{incl}xf_i(x,\mu^2)> -<\sigma_{incl}><xf_i(x,\mu^2)>}
{\Delta_{\sigma_{incl}} \Delta_{xf_i(x,\mu^2)}}
\label{eq:corrl}
\er
where $\rm {N_{rep}}$ is the number of replicas in the NNPDF sets,
$\rm {\sigma_{incl}}$ is the inclusive Z boson production cross section 
computed using MCFM and
$f_i(x,\mu^2)$ is the PDF for a given parton $i$ and the factorization scale
$\mu^2$. 
In the denominator, the $\Delta_{\sigma_{incl}}$ and
$\Delta_{xf_i(x,\mu^2)}$ are the
standard deviations of cross sections for $N_{rep}$ replicas of PDF sets and
the PDF replicas themselves respectively.

\begin{figure}[ht]
  \centering
  \includegraphics[width=0.45\textwidth]{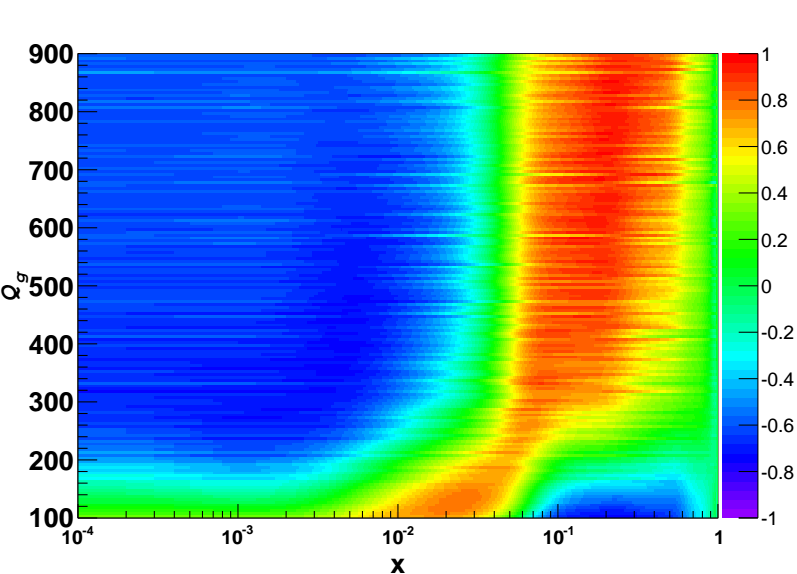}
  \includegraphics[width=0.45\textwidth]{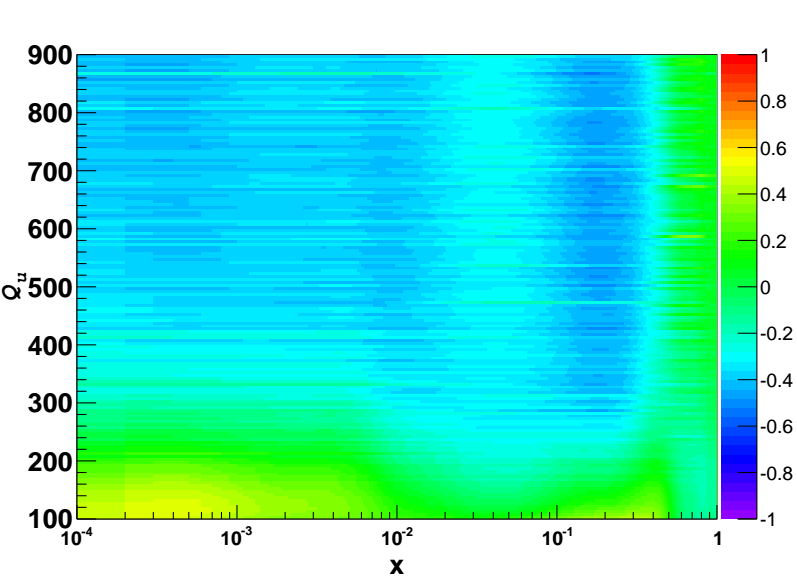}
  \includegraphics[width=0.45\textwidth]{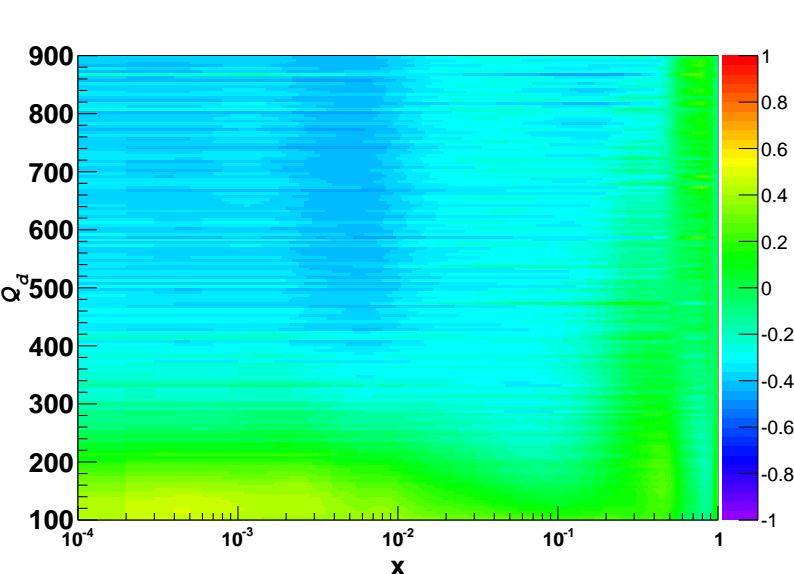}
  \includegraphics[width=0.45\textwidth]{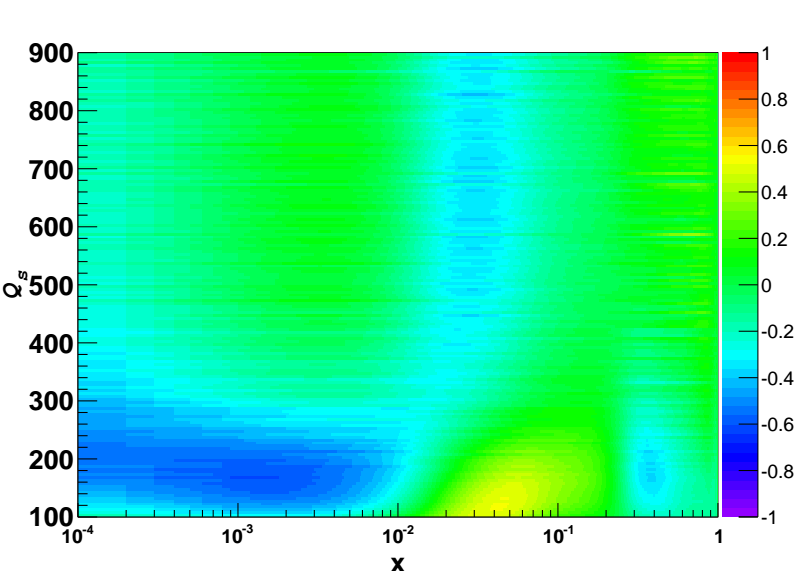}
  \caption{Correlation co-efficients(${\rm Q}_i$; $i$=g,u,d,s) between 
the gluon (top-left), the up
quark (top right), the
down quark (bottom left) and the strange quark (bottom-right) PDFs
and the inclusive Z boson production cross section for the 
$\ptz>40$~GeV range.
}
\label{fig:corrl}
\end{figure}

Fig.~\ref{fig:corrl} presents the correlation coefficients(${\rm Q}_i$) 
for the gluon, up, down and the strange quark PDFs in the 
$x$-${\rm Q}_i$ plane.
The correlation
co-efficient ${\rm Q}_i$ close to zero indicates that there is 
no correlation at all 
between the respective incoming parton and the production cross section.
Similarly, higher values of ${\rm Q}_i$ means presence of a strong 
correlation of the corresponding parton and cross sections, 
where as negative values mean anti-correlation.
The figure representing the gluon correlation
(upper left panel) indicates that the inclusive Z boson cross section
is strongly sensitive to gluon PDFs for the values of $x \sim 0.01 - 0.1$.

\section{The QCD Analysis}
\label{anal}
The NLO QCD analysis is performed using the framework of the open source code
{\tt HERAFitter(v1.1.0)}, the detailed description of 
which can be found in Ref.\cite{Alekhin:2014irh}.
In this analysis, the initial parametrization for the PDFs is assumed at the 
starting scale $\textbf{of}$ $Q^2_{\rm 0}= 1.9 $ GeV$^2$ and 
evolved to higher scales 
with the DGLAP~\cite{Gribov:1972ri,Lipatov:1974qm,Altarelli:1977zs,
Dokshitzer:1977sg} equations using {\tt QCDNUM}~\cite{Botje:2010ay}.
The combined data sets from the DIS neutral 
current (NC) and the charged current (CC) in $e^+p$ and $e^- p$ 
scattering at the 
H1 and ZEUS experiments as well as the CMS W muon charge asymmetry data 
are used in this study along with the data corresponding to
the measurement of inclusive Z production, as described above. 
In order to exploit the precise CMS lepton charge asymmetry data 
which are used to improve the constraints of the PDFs of 
light quarks~\cite{Chatrchyan:2013mza}, the CMS measurement of W 
charge asymmetry is added as an additional  
input along with the inclusive $\ptz$ data.     

At the starting scale, $Q^2_{\rm 0}$, the PDFs are parametrized 
using a generic form:
\br
xf(x) = A x^B (1-x)^C (1+Dx+Ex^2).
\er
Here $A$ is the normalization term and the behavior of the PDFs 
for low (high) values of Bjorken $x$ is regulated by the $B$ ($C$) term. 
The optimal parametrization for the PDF fit is
found through a parametrization scan as described in~\cite{Aaron:2009aa}: 
In the beginning, the scan is performed starting from a parametrization with 
a basic polynomial form and then additional parameters are allowed to vary, 
one parameter at a time. This scanning process continues till 
the reduction in $\chi^2$ reached to a value less than unity.

The final parametrized form with 14 free parameters for the five PDFs,
valence light quarks $\rm {xu_{v}(x)}, \rm{xd_v(x)}$, the 
anti quark 
$\rm {x\bar U(x), x\bar D(x)}$, where
$\rm{ x\bar U(x)=x\bar u(x)}$, $\rm{x\bar D(x)=x\bar d(x) + x \bar s(x)}$
and the gluon xg(x) is defined as:
\br
xg(x) &=& A_g x^{{B_g}}(1-x)^{C_g} + A^\prime_g x^{{B^\prime_g}}
(1-x)^{C^\prime_g}
 \nonumber \\
xu_{\rm v}(x) &=& A_{u_{\rm v}} x^{B_{u_{\rm v}}}(1-x)^{C_{u_{\rm v}}}
(1+E_{u_{\rm v}} x^2)
\nonumber \\
xd_{\rm v}(x) &=& A_{d_{\rm v}}x^{B_{d_{\rm v}}} (1-x)^{C_{d_{\rm v}}}\nn \\
x \bar U(x) &=& A_{\bar U} x^{B_{\bar U}}(1-x)^{C_{\bar U}}
\nonumber \\
x \bar D(x)&=& A_{\bar D} x^{B_{\bar D}}(1 -x)^{C_{\bar D}}
\label{eq:paramf}
\er
The normalization parameters 
$A_{u_{\rm v}}, A_{d_{\rm v}}$ and $A_g$ are constrained 
by the QCD sum rules. 
A more expanded form for g(x) is used with the choice $C^\prime_g=$25
following the approach of the MSTW group~\cite{Thorne:2006qt}.
The strange quark relation to $\bar D$ is defined as
\br
x\bar s = f_s x \bar D
\er
where $f_s$ is the fraction of strange quarks, 
$f_s= \frac{\bar s}{\bar d + \bar s} = 0.31\pm 0.08$~\cite{Martin:2009iq}.
Additional constraints applied are  $B_{\bar U}=B_{\bar D}$ and
$A_{\bar U} = A_{\bar D}(1-f_s)$.

The sources of experimental uncertainties in the measurement of 
the Z boson cross section are discussed in detail 
in the Ref.\cite{Khachatryan:2015oaa}
and taken into account in the fit through a covariance matrix.
The uncertainty due to the choice
of parametrization given by the Eq.\ref{eq:paramf} is evaluated by
assuming an alternate parametrization.
The parametrization uncertainties are estimated by including additional 
terms one by one in the polynomial expansion of Eq.3 for all parton 
densities following the procedure described in the Ref.~\cite{Aaron:2009aa}. 
The variation in the starting scale $Q^2_0$, is regarded as a 
parametrization uncertainty and is estimated by varying it within the range
$ 1.5 (f_s=0.29)\ge Q^2_0\ge 2.5 (f_s=0.34, m_c=1.6)$.
The parametrization uncertainty is constructed 
as an envelope built from the maximal differences between the PDFs.
The model uncertainties in the QCD fit are evaluated
by varying 
heavy quark masses $m_c$ and $m_b$, strange quark fraction parameter
$f_s$, and $Q^2_{min}$. 
In order to obtain these uncertainties, the model parameters
are varied between its maximum and minimum values one at a time 
in the fit.
The change of the fit due to this variation with respect to the 
central fit obtained using the nominal 
value of that parameter is estimated to be the uncertainty corresponding to
that model parameter. 
The model uncertainties with variations are presented 
in Table~\ref{tableModel}.
The experimental, model and parametrization uncertainties are
added in quadrature to obtain the total systematic uncertainty.
\\
\begin{table}[!ht]
\begin{center}
\caption{Maximum and minimum values of model parameters along with the nominal
value for the central fit.}
\begin{tabular}{cccc}
\hline
\hline
Parameters & Nominal value & Lower limit & Upper limit  \\
\hline
$f_S$      & 0.31   & 0.23             & 0.38 \\
$m_c$[GeV] & 1.4    & 1.35($Q_0^2$=1.8) & 1.65 \\
$m_b$[GeV] & 4.75   & 4.3              & 5.0 \\
$Q^2_{\rm min}$[GeV$^2$] & 3.5 &   2.5  & 5.0 \\
$Q^2_{0}$[GeV$^2$] & 1.9 & 1.5($f_s = 0.29)$)&2.5($f_S = 0.34, m_c=1.6~GeV$) \\
\hline
\end{tabular}
\label{tableModel}
\end{center}
\end{table}
\section{Results}
\begin{table}
\begin{center}
\caption{Partial $\chi^2$ per data point
and global $\chi^2$ per degrees of
freedom(dof) for the data sets used in 14-parameter fitting.}
\begin{tabular}{cccc}
\hline
\hline
Datasets & HERA I & HERA I +CMS(Asym) & HERA I + CMS full \\
\hline 
NC HERA H1-ZEUS $\rm {e^+p}$ & 109/145  & 109/145 & 109/145  \\
NC HERA H1-ZEUS $\rm {e^- p}$ & 400/379 & 401/379 & 411/379   \\
CC HERA H1-ZEUS $\rm {e^+p}$ & 19/34 & 19/34 & 19/34  \\
CC HERA H1-ZEUS $\rm {e^-p}$ & 27/34 & 30/34 & 31/34  \\
CMS W electron asymmetry & -- & 8.4/11 & 7.5/11 \\
CMS W muon asymmetry & -- & 13/11 & 13/11 \\
CMS Inclusive Z data & -- & -- & 78.5/40 \\
\hline
Total $\chi^2/dof$ & 555/578 & 580/600 & 668/640 \\
\hline
$\chi^2$ p value & 0.74 & 0.71 & 0.21 \\
\hline
\end{tabular}
\label{tableFits}
 \end{center}
\end{table}

\begin{figure}[ht]
  \centering
  \includegraphics[width=0.55\textwidth]{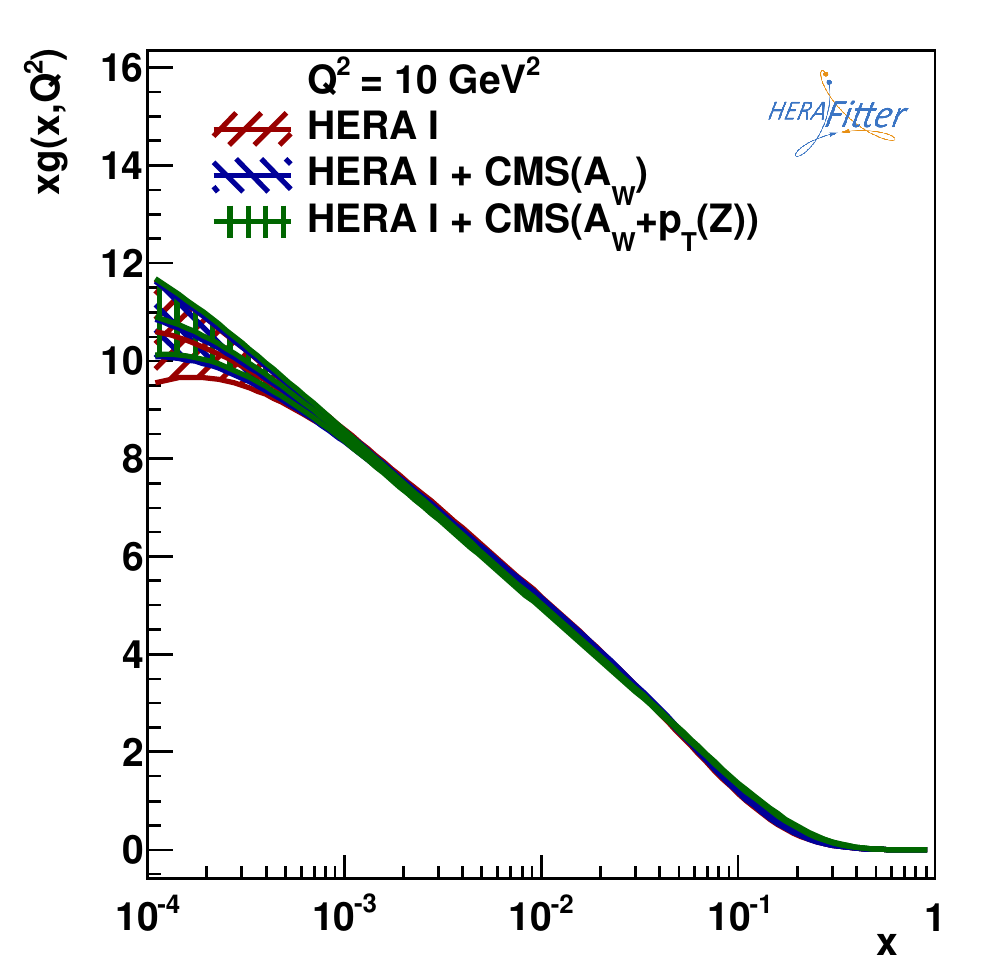}
  \includegraphics[width=0.55\textwidth]{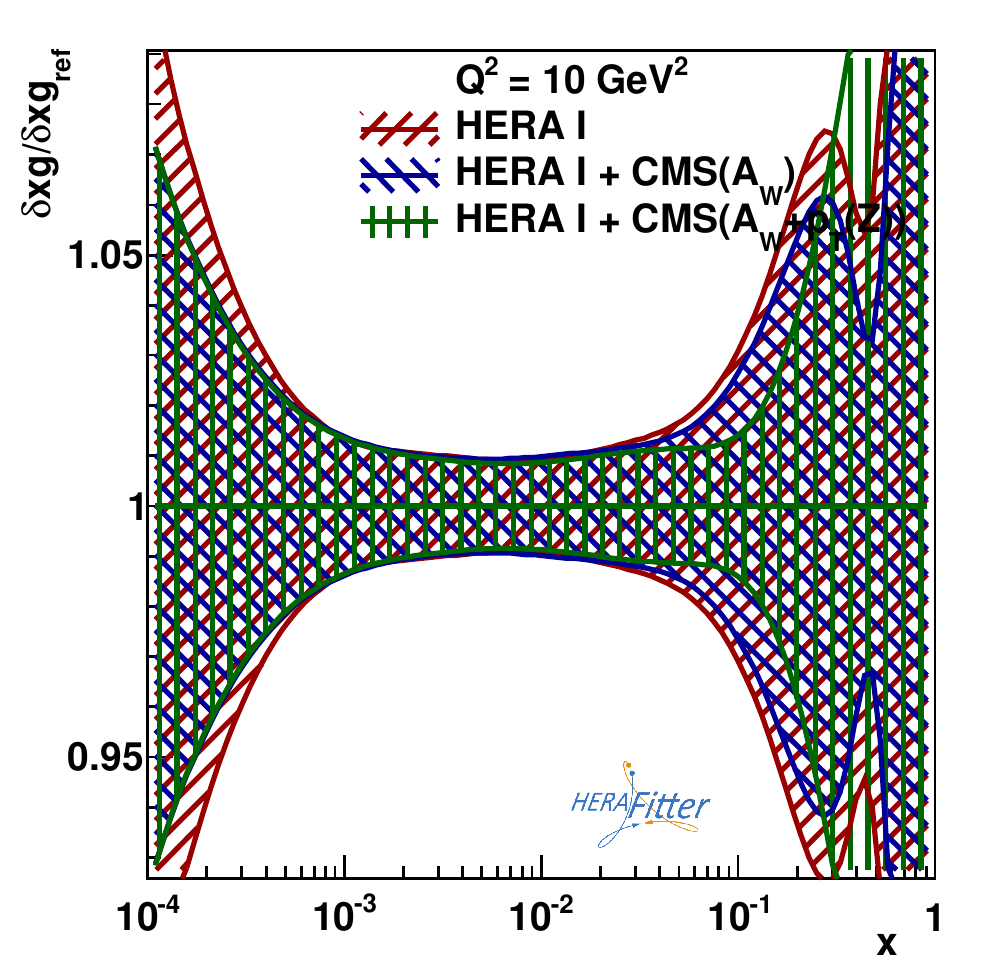}
\caption{ Distribution of gluon for $Q^2$=10 GeV$^2$. 
The bands correspond to experimental PDF uncertainties of the fit to 
HERA data only(red) and both
HERA and the CMS data(blue and green). 
}
\label{fig:gluon}
\end{figure}
The QCD analysis results are presented in Table~\ref{tableFits}.
The quality of the fit is found to be fairly good for the HERA-I data only and 
HERA-I with  CMS W charge asymmetry data. The combined fit of 
HERA-I data with CMS W and Z production data is observed to be 
higher due
to the fact that the disagreement
between the Z-boson
data and the corresponding MCFM theory prediction is relatively large.
However, the total $\chi^2/{\rm dof}$ is found to be reasonable.

\begin{figure}[ht]
  \centering
  \includegraphics[width=0.45\textwidth]{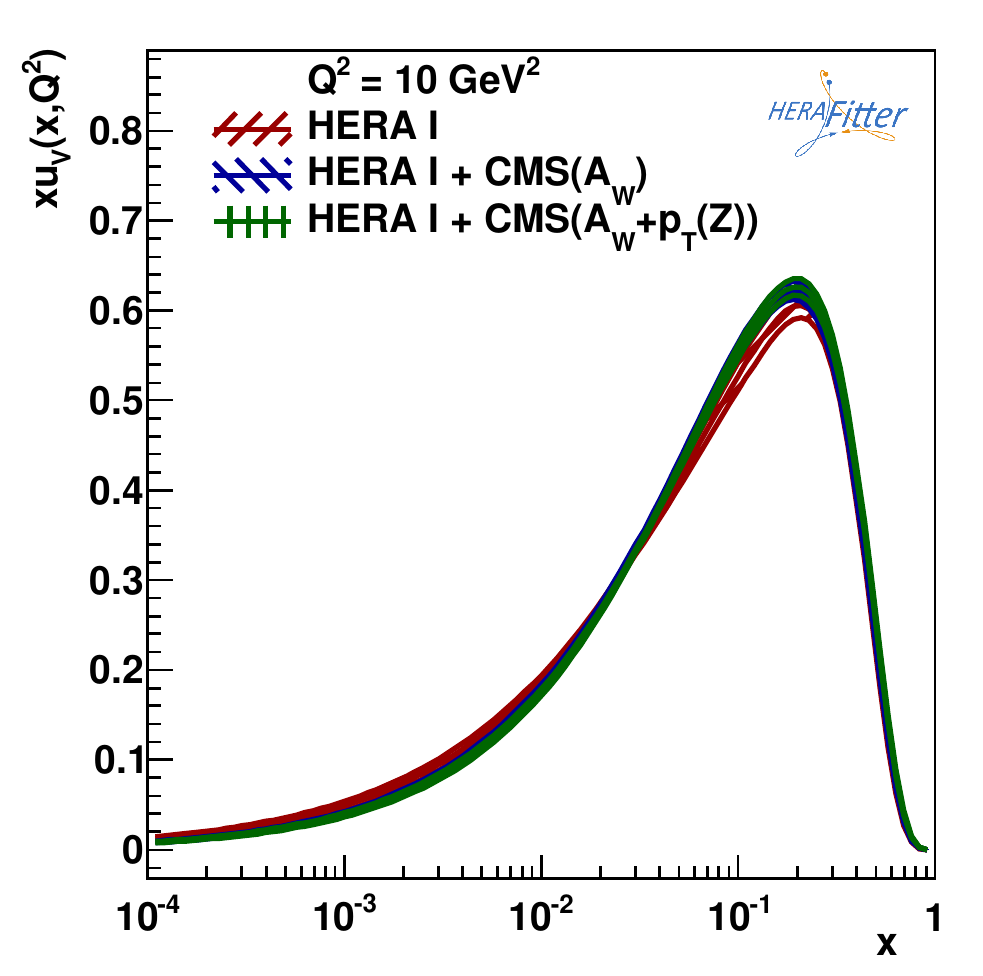}
  \includegraphics[width=0.45\textwidth]{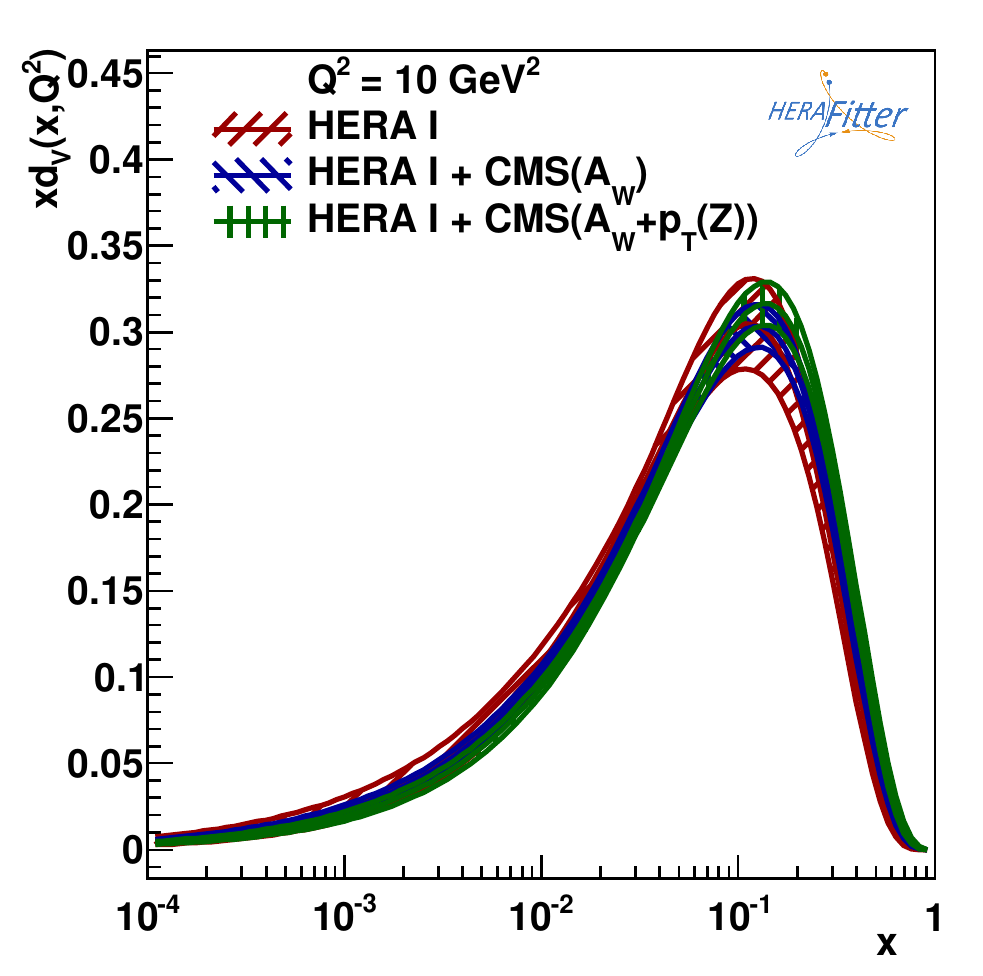}
  \includegraphics[width=0.45\textwidth]{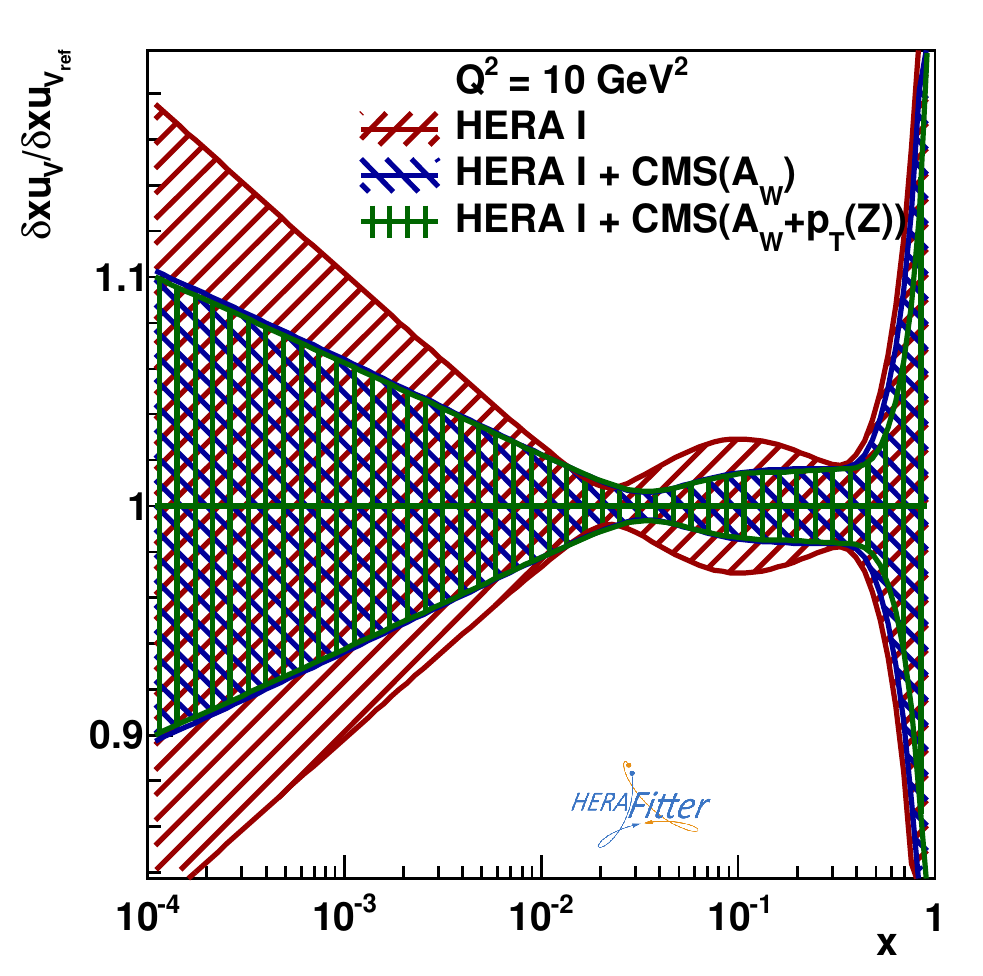}
  \includegraphics[width=0.45\textwidth]{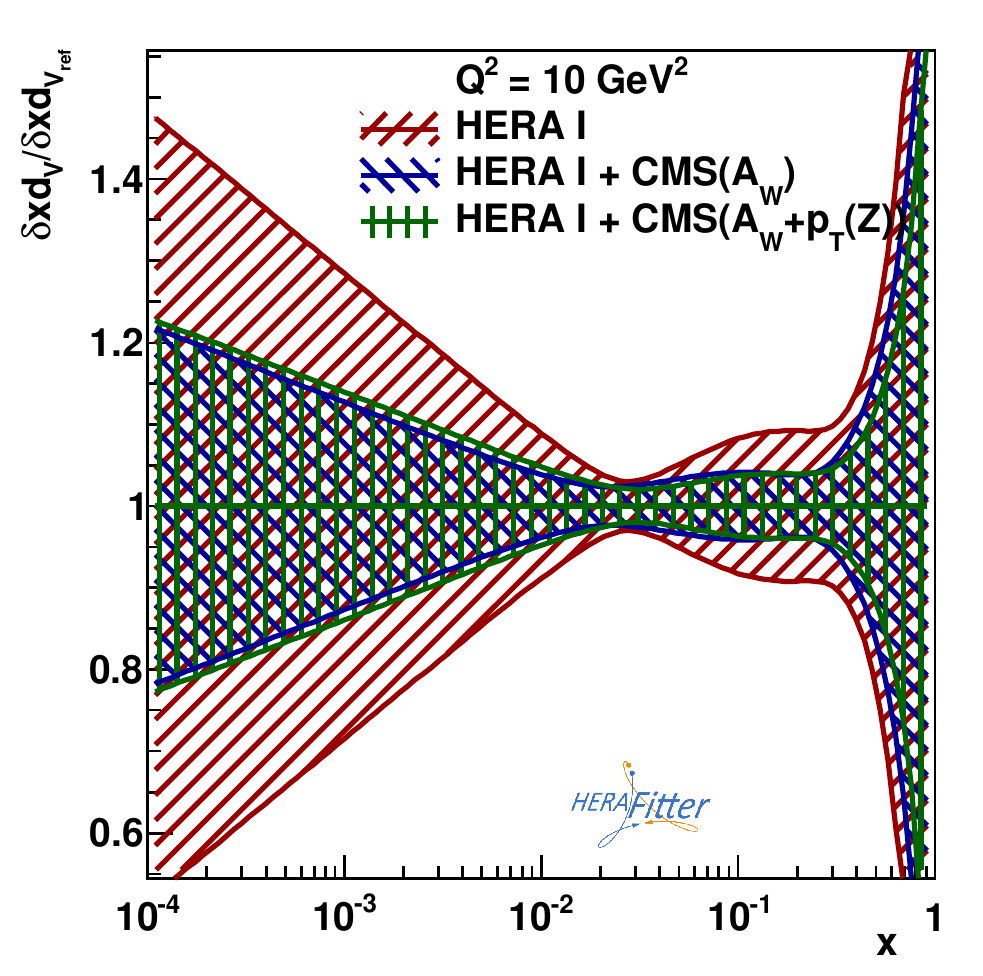}
\caption{ 
Same as in Fig.3, but for for u(left) and d(right) valence quarks.}
\label{fig:uval}
\end{figure}

\begin{figure}[ht]
  \centering
  \includegraphics[width=0.45\textwidth]{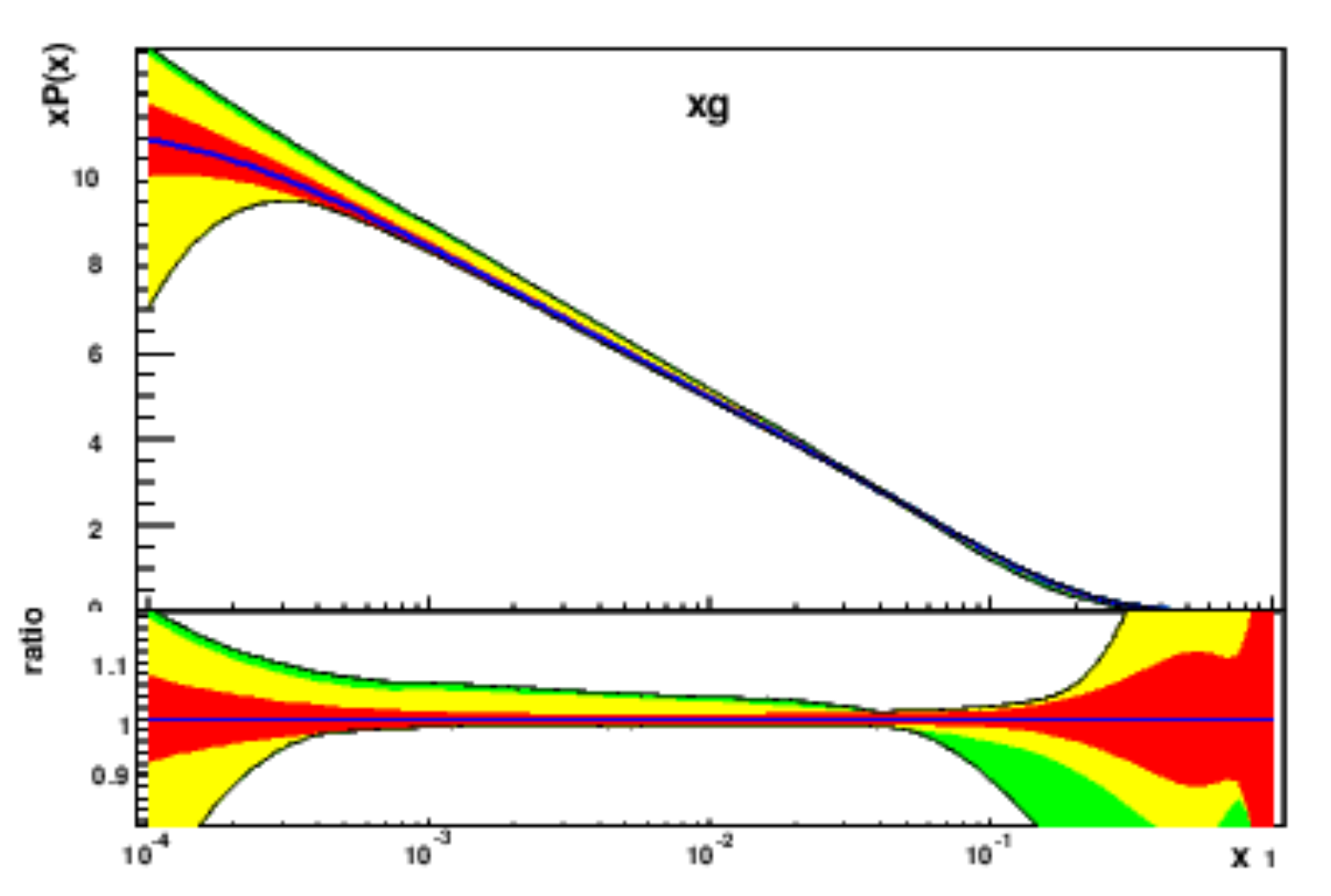}
  \includegraphics[width=0.45\textwidth]{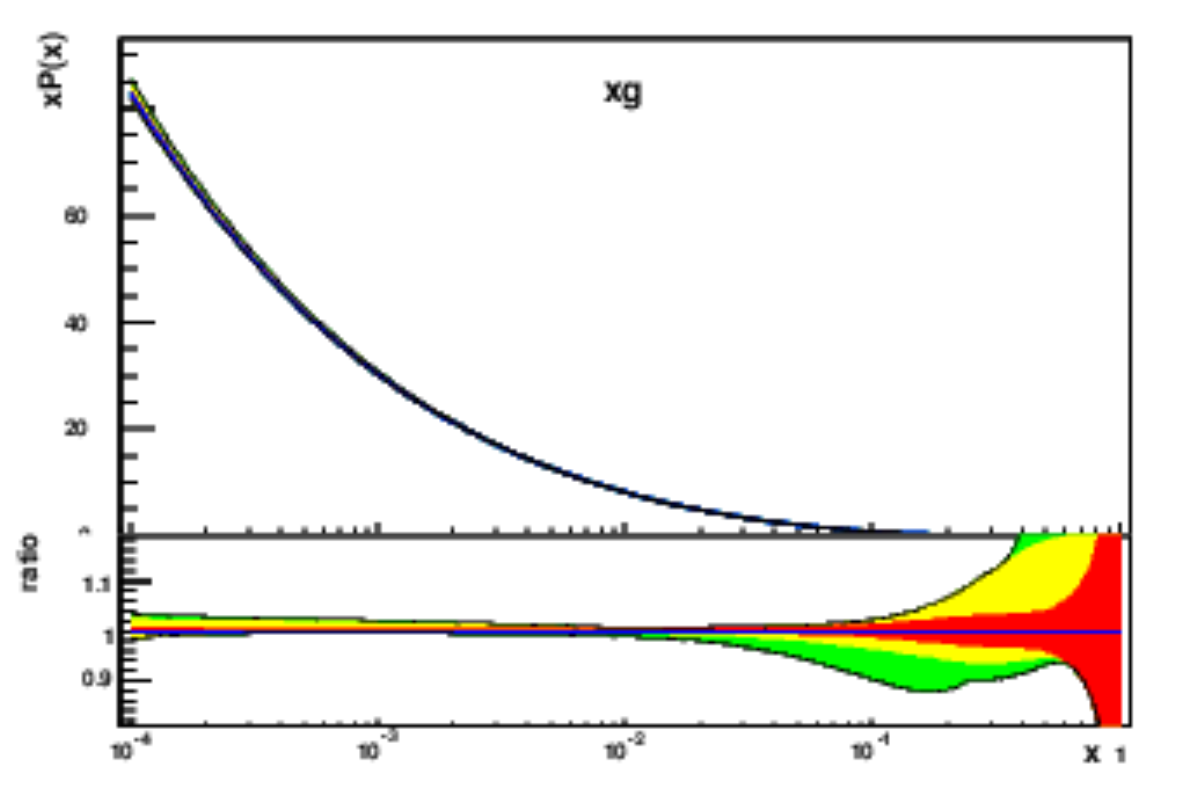}
  \includegraphics[width=0.45\textwidth]{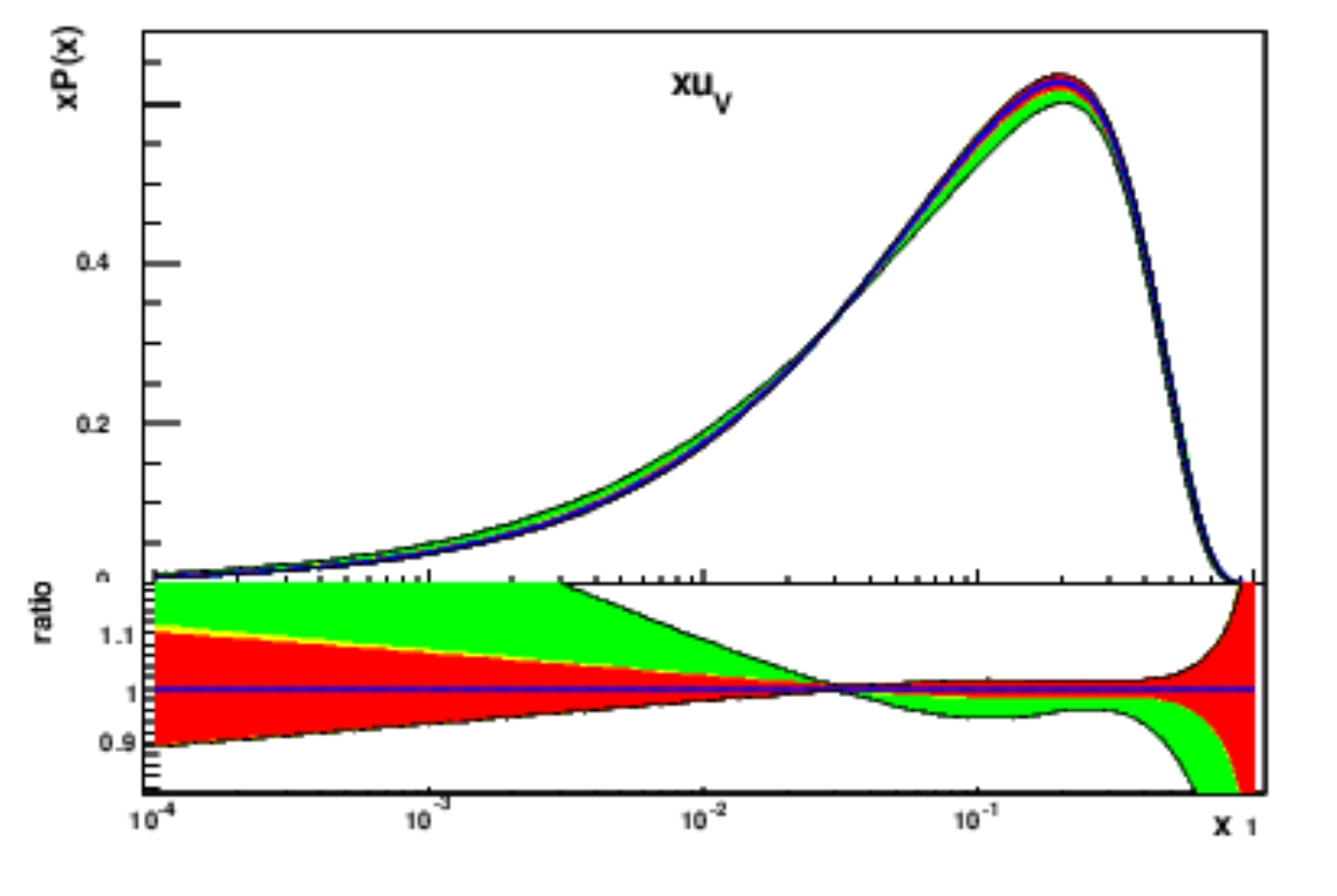}
  \includegraphics[width=0.45\textwidth]{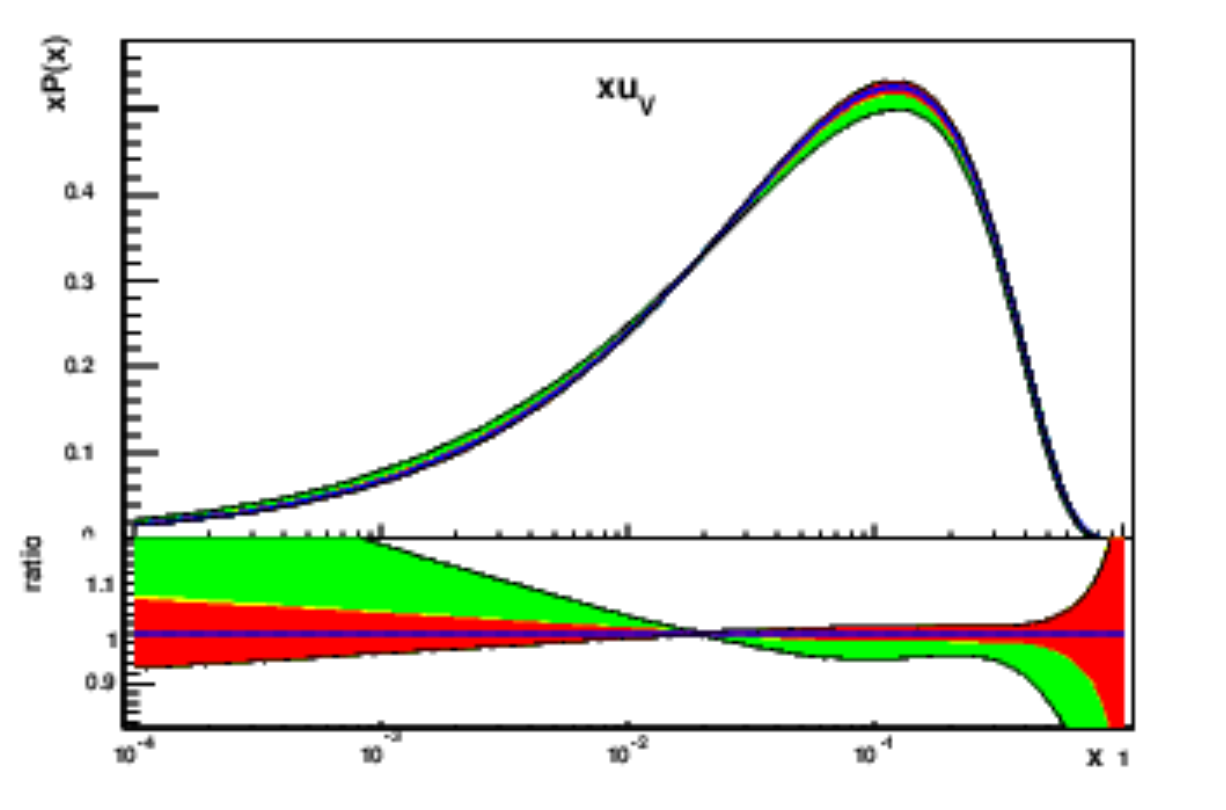}
 \includegraphics[width=0.45\textwidth]{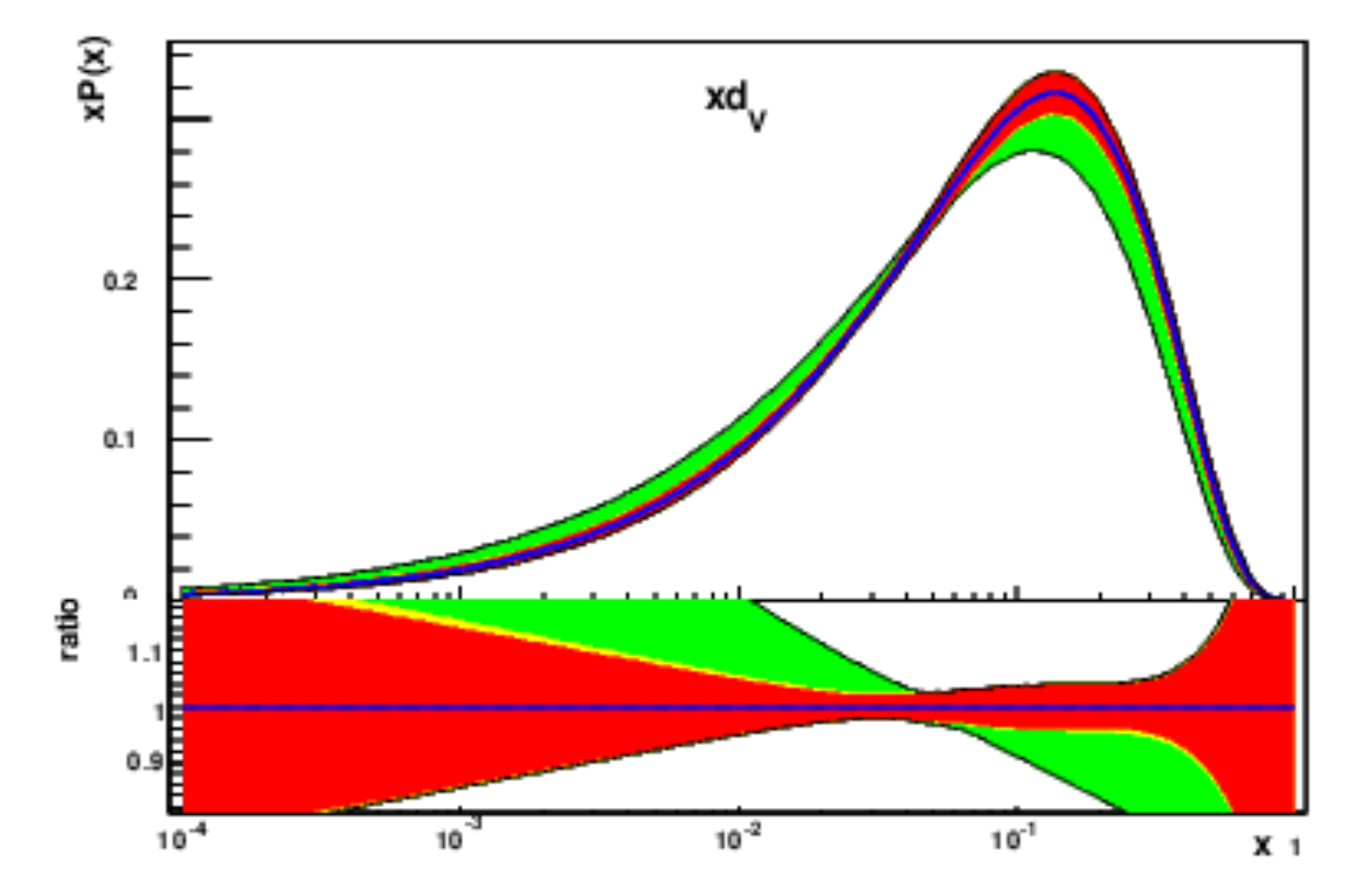}
  \includegraphics[width=0.45\textwidth]{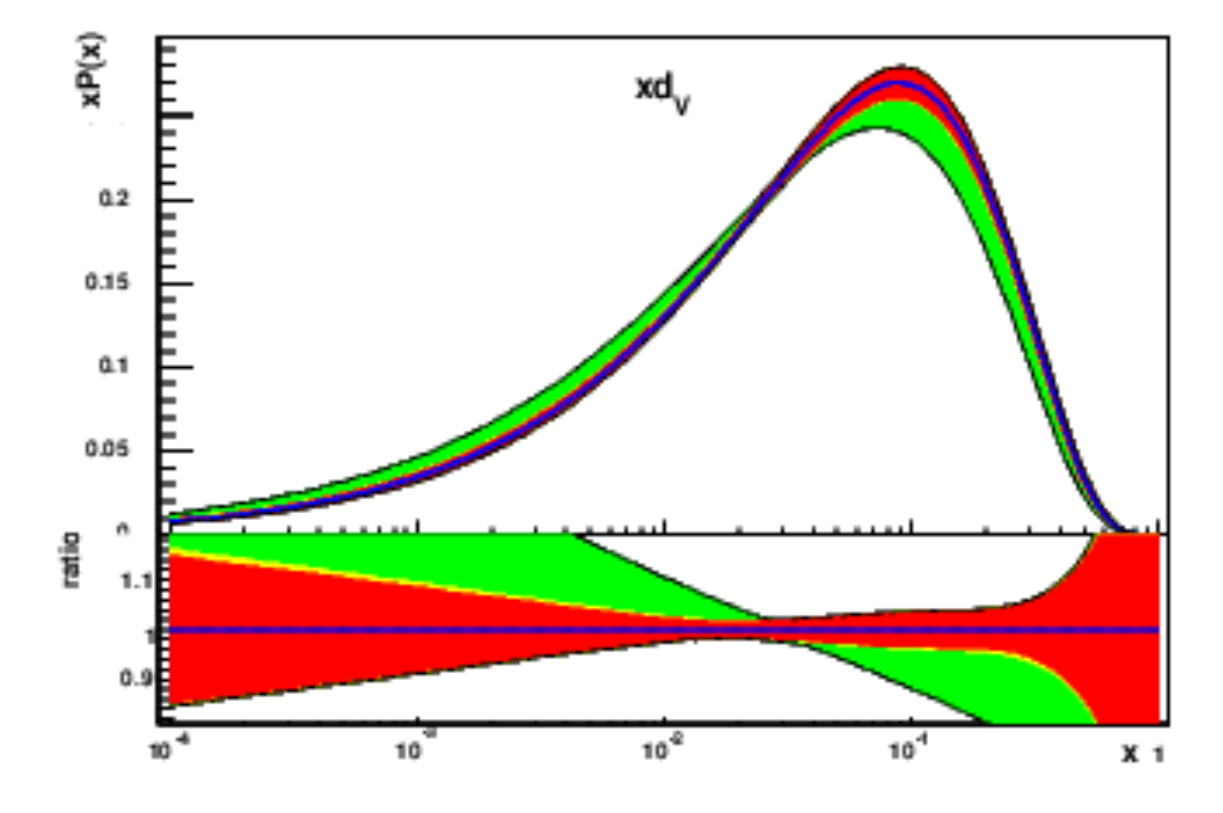}
\caption{Constrained parton density functions for gluon (top), u (middle) and
    d valence (bottom) quarks from the QCD analysis of CMS Z-boson 
    data at the scales $Q^2 = 10$~GeV$^2$
 (left panel) and $M_Z^2$ (right panel).
 The uncertainties include due to the experimental (red),
 the model (yellow) and the parametrization variation (green). 
 All uncertainties are added in quadrature.
}
\label{fig:pdfdist}
\end{figure}

The impact of the CMS data to PDFs is illustrated by comparing the PDF 
fits with
the HERA DIS data alone as shown for the gluon in Fig.~\ref{fig:gluon} and
u and d valence quarks in Fig.~\ref{fig:uval}. 
In order to understand the improvement in gluon distribution 
due to the addition of the CMS data, the 
ratio of the relative uncertainties of the fitted gluon distribution  
obtained using the combined datasets 
and the DIS data only is presented in the same figure at the bottom panel.
A change of shape of the gluon distribution due to the 
inclusion of the CMS Z boson data is visible at low $x$,
in particular around the value of $x \sim 0.1$ 
where a strong correlation of the gluon PDF with the Z boson production 
cross section is observed, as shown in
Fig.~\ref{fig:corrl}.
In addition, the Fig.~\ref{fig:uval} shows that the CMS 
W charge asymmetry is more sensitive to
the light quark distributions.
The constrained NLO distributions of the gluon, 
u and d valence quarks are shown in Fig.~\ref{fig:pdfdist}
for two values of $Q^2=$10 GeV$^2$ 
and $M_Z^2$. 
In general, as observed, the uncertainty due to the parametrization 
is the dominant one.

\section{Summary}
The sensitivity of the CMS Z boson production measurement to PDFs 
at $\sqrt{s}=8$~TeV is reported in this letter.
The theory predictions corresponding to the CMS Z boson production 
measurement are obtained
from the MCFM based calculations at NLO. 
The studies of the initial parton correlations with the Z boson cross 
section indicate the sensitivity of the gluon distribution in this process.
A comparison between the measured Z boson cross section in various 
$\ptz$ and $\rm Y(Z)$ bins and the corresponding MCFM based theory predictions 
at NLO shows an agreement at the level of $\sim$10\%, 
however the shapes of the distributions of 
both the data and the theory agree reasonably well. 
Similar level of agreement is also observed by the calculations performed 
with FEWZ. As evaluated with the MCFM based theory predictions, 
the uncertainty due to the QCD scales is found to be 
the most dominant of the order of 5-7\%.
It is to be noted that the inclusive double differential Z boson 
cross section is measured with an overall precision of about 3-4\% which is 
remarkably precise for any measurement at the hadron colliders. 

The NLO QCD analysis is performed within the framework of the 
{\tt HERAFitter}, 
fitting the CMS Z boson production and the W asymmetry measurements together 
with the HERA-I DIS charged and neutral current data.
The results of this QCD analysis indicate the 
improvement($\sim$5-7\%) in the gluon PDFs 
around the region of $x \sim$ 0.1. 
The current analysis demonstrates the limited constraints 
on the gluon PDFs using inclusive Z boson data. Therefore, in order to 
describe the very precisely 
measured Z boson cross section, more accurate 
theoretical predictions accessible via the fast techniques 
to PDF fits are needed
which are expected to reduce the level of disagreement between 
the data and theory.   

\bibliographystyle{utphys}
\bibliography{pdfpaper}{}
\end{document}